\documentclass[final]{planetary}

\title{How long could volcanic plumes persist in the Venus atmosphere?}
\author[1,*]{Maureen Cohen\,\orcidlink{0000-0001-5014-4174}}
\author[1]{James Holmes\,\orcidlink{0000-0003-3018-2135}}
\author[1]{Joanna Egan\, \orcidlink{0000-0001-9676-3990}}
\author[1]{Stephen Lewis\, \orcidlink{0000-0001-7237-6494}}
\author[1]{Manish R. Patel\, \orcidlink{0000-0002-8223-3566}}
\affil[1]{The Open University, Milton Keynes, United Kingdom, https://ror.org/05mzfcs16}
\affil[*]{\textit{Corresponding author:} Maureen Cohen — \texttt{maureen.cohen@open.ac.uk}}

\AtEveryBibitem{%
    \clearfield{note}%
    \clearfield{eprint}%
    \clearfield{urlyear}%
    \clearfield{urlmonth}%
    \clearfield{urlday}
    }
\DeclareUnicodeCharacter{03BC}{\textmu}
\DeclareUnicodeCharacter{00B5}{\textmu}
\date{\small Submission date: \today}

\begin{document}

\maketitle
\thispagestyle{firstpage}

\mstype{Article}

\summary{This study determines volcanic plume dispersal times in the Venus atmosphere using the 3-D Venus Planetary Climate Model. Results indicate that near-surface plumes of inert gases persist for days and could be detectable, while tropospheric plumes can be masked by atmospheric wave activity.}

\begin{abstract}
Resolving the ambiguity surrounding current Venusian volcanism is a primary objective for upcoming missions like EnVision and VERITAS. This study uses the Venus Planetary Climate Model to simulate the dispersal and detectability of volcanic plumes containing water vapour (H$_{2}$O), hydrogen chloride (HCl), carbon monoxide (CO), and carbonyl sulphide (OCS) within a three-dimensional atmospheric environment. We model localised gas enhancements at altitudes probed by nightside spectral windows (8.62, 20.91, and 35.45 km above surface). Results indicate that plumes persist longest in the deep atmosphere (8.62 km) at equatorial latitudes, where H$_{2}$O enhancements remain distinguishable from low background variability for up to 100 hours, forming distinct downstream streaks. In contrast, at 35 km, high intrinsic variability in chemically active species (CO and OCS) driven by atmospheric dynamics obscures plume signatures and chemically inert species disperse within 0.5-2 Earth days. Plumes at high latitudes disperse more rapidly than at low latitudes at all altitudes. We perform sensitivity tests with trace gas increases of 50\%, 40\%, 30\%, 20\%, 10\%, and 3\% above baseline and plume injection times of 6, 12, 28, 38, and 126 hours at all three altitudes and two latitudes. We conclude that observing inert gases, in particular H$_{2}$O, at low altitudes and latitudes offers the best opportunity for successful plume detection.
\end{abstract}

\vfill
\noindent
\begin{tabular}{p{0.01\textwidth} p{0.32\textwidth} p{0.01\textwidth} p{0.43\textwidth} p{0.1\textwidth}}
\cline{1-5}
\scriptsize{\textcopyright}
& \scriptsize{The Author(s) 2026.}
&
& \multirow{2}{*}{\parbox{0.43\textwidth}{\vspace{1mm}\scriptsize{This work is licensed under a \href{https://creativecommons.org/licenses/by/4.0/}{Creative Commons Attribution License 4.0} which permits unrestricted use, distribution, and reproduction in any medium, provided the original work is properly cited.}}}
& \vspace{1mm}\includegraphics[height=1.5em]{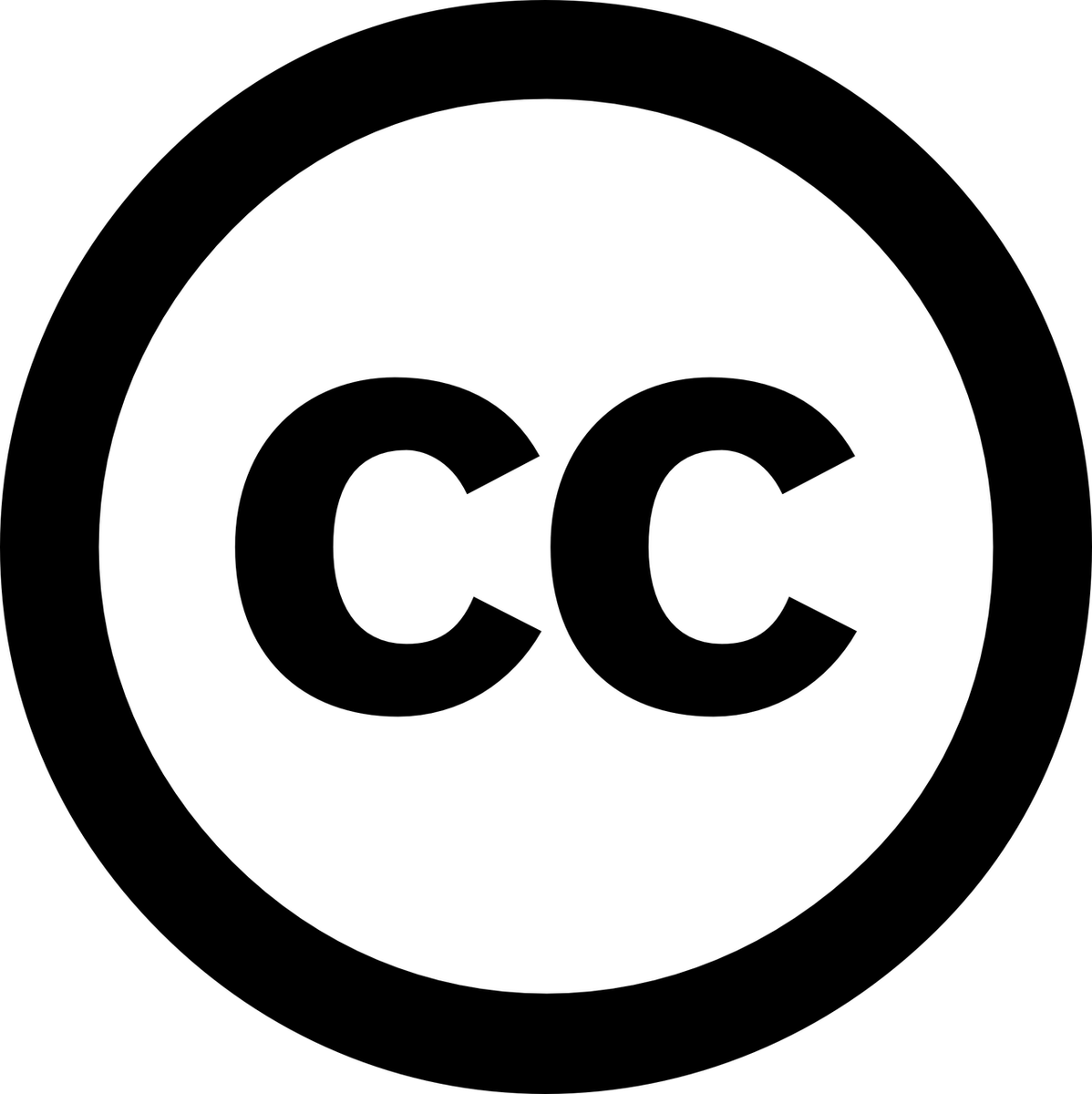}\hspace{0.15em}\includegraphics[height=1.5em]{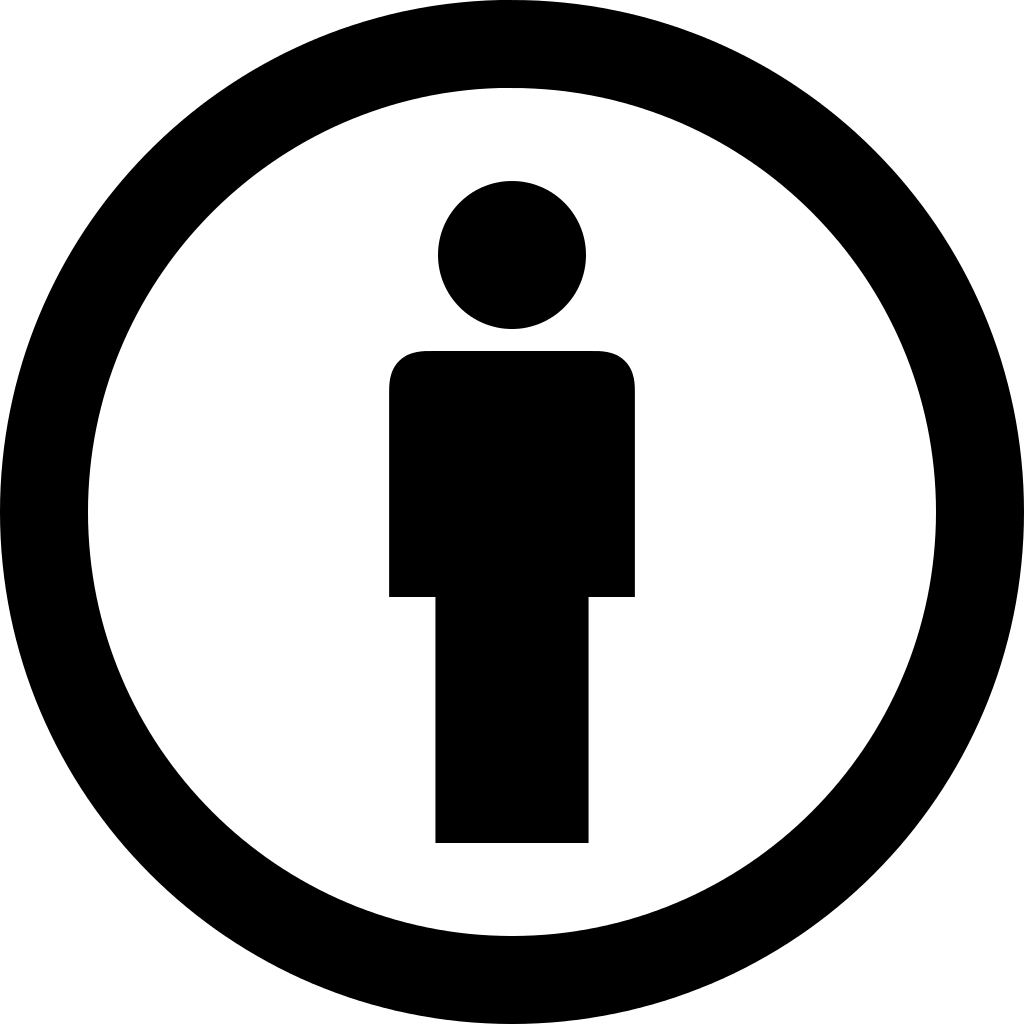}
\end{tabular}

\newpage

\section{Introduction} \label{sec:intro}
The surface of Venus is covered by extensive volcanic features, including lava plains, shield volcanoes, and arachnoid structures, but the question of how volcanically active Venus is today has yet to be definitively answered \parencites{filiberto_assessing_2025, king_volcanic_2022, ivanov_chapter_2015, ivanov_history_2013}. Evidence of recent volcanic activity includes possible changes in the shape of a lava vent \parencite{herrick_surface_2023} and in surface radar backscatter \parencite{sulcanese_evidence_2024} during the Magellan mission, laboratory studies of weathering rates of surface minerals \parencite{filiberto_present-day_2020}, temporary bright spots in Ganiki Chasma that could correlate with lava flows \parencite{shalygin_active_2015}, and thermal emissivity "hot spots" in nine locations \parencite{smrekar_recent_2010}. Some estimates of the frequency of volcanic eruptions have proposed a range of 42 to 120 eruptions per year \parencites{van_zelst_comment_2022, byrne_estimates_2022}. Understanding the current level of geologic activity is essential for constraining the planet's interior dynamics and climate evolution, ultimately shedding light on why and how the environments of Earth and Venus diverged. 

Resolving the ambiguity surrounding Venusian volcanism is a primary science objective for the next generation of planetary missions. The European Space Agency's (ESA) EnVision mission \parencites{kiefer_envision_2023, widemann_venus_2023} and NASA's VERITAS mission, both scheduled for launch in the early 2030s, are designed with specific capabilities to detect active surface and atmospheric processes. A key strategy for these missions involves targeted observations of the Venus atmosphere in search of plumes from volcanic eruptions. The EnVision mission will search for plumes beneath the planet's cloud decks using its VenSpec-H and VenSpec-M instruments \parencites{cock_design_2025, neefs_venspec-h_2025, helbert_venspec_2019}, as well as volcanic gases above the clouds using the VenSpec-U instrument. Similarly, the VERITAS mission will use the Venus Emissivity Mapper (VEM) to scan for anomalies, with a particular capability to search for increases in water vapour near the surface \parencite{pertenais_venus_2025}. 

The reasoning behind this observing strategy is that an active eruption will inject a localised plume of hot, chemically distinct gas into the atmosphere, creating a detectable contrast against the background. Observers have long exploited a number of near-infrared windows in the spectrum of Venus through which radiance can escape from beneath the clouds on the planet's nightside hemisphere \parencites{cardesin-moinelo_global_2020, tsang_general_2017, arney_spatially_2014, cotton_distribution_2012, wilson_analysis_2009, tsang_tropospheric_2008, marcq_latitudinal_2005, crisp_dark_1991, meadows_ground-based_1996, bezard_deep_1990, crisp_nature_1989}. Several gas species of potentially volcanic origin can be detected in these spectral windows, and recent modeling of volcanic gas plumes suggests that such injections by explosive volcanism could indeed have a measurable temporary effect on the planet's spectrum \parencite{dias_volcanic_2025, muller_detecting_2025}. Specifically, \textcite{dias_volcanic_2025} identifies enhancements in water vapour, carbon monoxide, and carbonyl sulphide in the deep atmosphere (30–40 km) as having the most significant potential impact on observed radiance. 

Interpreting such changes in radiance as volcanic plumes is complicated by conflicting theories regarding the behaviour of plumes and the drivers of atmospheric variability. A central point of contention is the maximum altitude to which a buoyant volcanic plume can rise on Venus, as the statically stable, high-pressure atmosphere of Venus presents a barrier to explosive volcanism in most scenarios \parencite{airey_explosive_2015}. Several studies have used simulations of buoyant volcanic plumes under different conditions (e.g. vent size, crustal volatile content, magmatic temperature) to investigate how high a plume could rise, and in particular whether it could extend to the cloud tops \parencites{lefevre_eruption_2025, glaze_explosive_2011, glaze_transport_1999, thornhill_theoretical_1993}. For example, recent work by \textcite{lefevre_eruption_2025} found that 87\% of modelled plumes did not rise above 15 km, approximately 2.5\% could rise to the cloud bottom at 47 km, and none reached the cloud tops. While these theoretical studies do allow plumes to rise to the altitudes of the nightside spectral windows, the simulations are one-dimensional and include limited representation of the surrounding atmosphere. This means they do not capture plume behaviour in the realistic, three-dimensional atmospheric environment of Venus. In particular, plumes could dissipate quickly in the high wind speeds of Venus and could be indistinguishable from the background variability in trace gas species \parencite{wilson_possible_2024}. 

Our work aims to close this gap by representing volcanic plumes as localised enhancements of trace gases in a fully three-dimensional Venus coupled photochemistry-climate model, the Venus Planetary Climate Model \parencite{stolzenbach_three-dimensional_2023, lebonnois_superrotation_2010}. We simulate plumes which increase the volume mixing ratio of the volcanic species available in the model by a range from 3\% to 50\% at altitudes observable in three nightside spectral windows. We then place upper limits on how long it takes for the concentrations to return to the baseline value (the plume dispersal time) and whether the enhancement is detectable above the background variability in each trace gas species due to non-volcanic chemical and dynamical processes. Our study does not address whether or not explosive eruptions can occur from a geophysical perspective or how high buoyant plumes can ascend; rather, we test whether the atmospheric dynamics and chemistry would allow such plumes to be detectable if they do exist. 

We describe the Venus Planetary Climate Model and our approach to simulating plumes in Section \ref{sec:methods}. The results of our simulations are presented in Section \ref{sec:results}. In Section \ref{sec:discussion}, we discuss the implications of our study by identifying the best observing prospects for tropospheric plumes in terms of altitude, latitude, and species. We conclude in Section \ref{sec:conclusion}.

\section{Methods} \label{sec:methods}
\subsection{Model} \label{subsec:methods_model}
We conduct our study using the Venus Planetary Climate Model (VPCM). The VPCM is a state-of-the-art, open-source forward model of the Venus atmosphere originally described in \textcite{lebonnois_superrotation_2010} and used in numerous subsequent studies \parencites{lai_planetary-scale_2024,lefevre_impact_2022,lefevre_mesoscale_2020,garate-lopez_latitudinal_2018,lebonnois_planetary_2018,navarro_atmospheric_2018,lebonnois_wave_2016,marcq_simulations_2013}. The VPCM consists of a dynamical core \parencite{hourdin_lmdz4_2006} that calculates the horizontal and vertical wind fields, pressure, potential temperature, and tracer abundances on a longitude-latitude grid and a hybrid sigma-pressure vertical coordinate. The core is coupled to Venus-specific physics routines, including a full radiative transfer scheme and a specific heat capacity of the atmosphere that varies with altitude \parencite{lebonnois_superrotation_2010}. \textcite{cohen_planetary_2024} added a diagnostic age-of-air scheme which uses the VPCM's tracer transport routines to calculate transport timescales of a passive gas-phase tracer from a source region to any other location in the simulated atmosphere. The spatial grid used in our study has a horizontal resolution of 3.75\textdegree x1.875\textdegree~(96 longitudes by 96 latitudes) and a vertical resolution of 78 layers unevenly spaced between the surface and 98 km altitude, with higher resolution near the surface. This leads to a gridbox size of around 400x200 km in the horizontal and a depth of around 1-2 km in the altitudes we study. The altitudes used in the figures in Section \ref{sec:results} below are determined by calculating the long-term, area-weighted global mean of the geopotential height. 

\textcite{stolzenbach_three-dimensional_2023} coupled the VPCM's dynamics and physics to a photochemistry scheme incorporating 33 chemical species, 101 chemical reactions, and 20 photodissociation reactions. This chemistry scheme was expanded by \textcite{egan_is_2025}, \textcite{martinez_three-dimensional_2024}, and \textcite{streel_strong_2024} to incorporate more species and an ionosphere extension. As the ionosphere is too high to be affected by volcanic eruptions, we run the VPCM at its standard vertical extent of 98 km. The chemistry scheme includes reactions extending down to 35~km, but does not incorporate thermochemical reactions believed to occur in the lowest 10~km of the atmosphere and modelled by \textcite{krasnopolsky_s3_2013} and \textcite{krasnopolsky_chemical_2007}. These thermochemical reactions are excluded because their chemical timescales are typically orders of magnitude longer than the photochemical and cloud chemistry timescales \parencite{stolzenbach_three-dimensional_2023}, extending to hundreds of years (e.g. 270 years for OCS according to \textcite{krasnopolsky_chemical_2007}). As shown in Section \ref{sec:results}, the plumes in our study disperse on much shorter timescales.

The current chemistry scheme includes the following trace gas species which can be produced through volcanic activity: water vapour (H$_2$O), carbon monoxide (CO), carbonyl sulphide (OCS), hydrogen chloride (HCl), and sulphur dioxide (SO$_2$). We use H$_2$O, CO, OCS, and HCl as the constituents of the volcanic plumes in our simulations. \textcite{dias_volcanic_2025} quantified the impact of increases in H$_2$O, CO, and OCS on the 2.3 \textmu m thermal emission window at 30-40 km, where VenSpec-H will observe. While their study does not cover HCl, this gas is detectable on the nightside at 1.7 \textmu m \parencites{arney_spatially_2014, iwagami_hemispheric_2008} and could be emitted by eruptions. Likewise, H$_2$O has been observed on the nightside at 1.18 \textmu m \parencite{bezard_110_2011a, bezard_water_2009} and a science goal of VenSpec-M will be to detect water vapour from volcanic emissions near the surface \parencite{alemanno_mapping_2025}. We do not simulate SO$_2$ plumes as \textcite{dias_volcanic_2025} found even large increases are unlikely to be detectable on the nightside below the clouds. Although CO$_2$ has been found to be a likely constituent gas of volcanic eruptions \parencite{gaillard_theoretical_2014}, we do not model CO$_2$ plumes because this species is not treated as a trace gas in the VPCM and localised CO$_2$ enhancements are not observable.

The chemistry of the Venus atmosphere is not fully understood and the VPCM's photochemistry scheme does not match the observed abundances for all trace gases \parencite{stolzenbach_three-dimensional_2023}. The model is spun up from an initial condition in which the volcanic trace gases are initialised as follows: H$_2$O at 30 parts per million by volume (ppm), HCl at 0.4 ppm, OCS at 3 ppm, and CO at 25 ppm. The SO$_2$ abundance is fixed at 130 ppm in the atmosphere below the cloud decks and follows a prescribed exponential decline across the cloud decks to reach a value of 0.13 ppm at the cloud tops to match the observed profile. Above the cloud-top altitude (70 km), the SO$_2$ abundance is freely evolving. The simulation is run until all species reach a steady state. This steady-state output is then used as the initial condition for the plume simulations. The initial condition therefore already contains a fully developed superrotation, planetary-scale waves, the cloud-level Hadley cells, and other features of the general circulation. The H$_2$O and HCl abundances correspond well to the observed values of these species, but the CO and OCS abundances in the deep atmosphere (35 km) reach a global long-term mean of 7.35 ppm and 1 ppm, lower than the observed 25 ppm for CO and at the bottom of the observed range of 1-3 ppm for OCS. Where they differ, we take these global mean steady-state values, not the observed values, as the baseline to determine the intensity of the plumes and the dispersal times in the simulations. In other words, the scaling factor is applied to 7.35 ppm instead of 25 ppm for CO. This allows us to estimate the dispersal times and variability based on relative changes even though the model does not reproduce absolute abundances.

\subsection{Plume experiments} \label{subsec:methods_plume_exps}
We model the volcanic plumes as an enhancement of the volume mixing ratio (vmr) of the volcanic gas species in a single gridbox. This approach can be considered a representation of the umbrella phase of an eruption, after the plume has stopped ascending and spread out horizontally \parencites{sandvik_methodology_2021, burton_use_2020, sparks_dimensions_1986}. Only the largest Plinian-style eruptions on Earth have produced umbrellas with a radius similar in size to a gridbox of the VPCM (200-400 km). For example, the Mount Pinatubo eruption in 1991 produced an umbrella with a maximum radius of 570 km (\textcite{prata_radial_2025}, Table 2). In addition, we do not take into account any dispersal effects during stages of the eruption before the formation of the umbrella. This means the calculations presented in Section \ref{sec:results} should be taken as upper limits on the timescales for observing volcanic plumes. The plume composition at each altitude is determined by the gases observable in the corresponding spectral window rather than any physical constraint on distribution. For each species, we enhance the volume mixing ratio by a factor of 1.5 times the steady-state background value to align our study with the work of \textcite{dias_volcanic_2025}, who use similar values in their investigation of the spectral impact of plumes. The 1.5x scaling factor simulation is referred to as the nominal simulation throughout the remainder of this text. We further perform sensitivity tests with plumes consisting of H$_2$O only using scaling factors of 1.4, 1.3, 1.2, 1.1, and 1.03 times the background value to test the visibility of weaker eruptions.

Our nominal simulation contains the following six plumes:
\begin{table}[!ht]
    \centering
    \scriptsize
    \renewcommand{\arraystretch}{1.0}
        \begin{tabular}{c c c c}
            \toprule
            \textbf{Plume} & \textbf{Location} & \textbf{Composition (ppm)} & \textbf{Observability} \\
            \hline
            1. & 2\textdegree N, 168\textdegree W, 8.62 km & H$_2$O (45) & 1.18 \textmu m window  \\
            2. & 64\textdegree N, 4\textdegree E, 8.62 km & H$_2$O (45) & 1.18 \textmu m window \\
            3. & 2\textdegree N, 168\textdegree W, 20.91 km & H$_2$O (45), HCl (0.6) & 1.74 \textmu m window \\
            4. & 64\textdegree N, 4\textdegree E, 20.91 km & H$_2$O (45), HCl (0.6) & 1.74 \textmu m window \\
            5. & 2\textdegree N, 168\textdegree W, 35.45 km & H$_2$O (45), HCl (0.6), CO (11), OCS (1.5) & 2.3 \textmu m window \\
            6. & 64\textdegree N, 4\textdegree E, 35.45 km & H$_2$O (45), HCl (0.6), CO (11), OCS (1.5) & 2.3 \textmu m window \\
            \hline
        \end{tabular}
        \caption{Simulated plume locations and compositions, as well as the observability of each site.}
        \label{tab:plumes}
\end{table}

The sensitivity tests contain the same set of plumes in Table \ref{tab:plumes} repeated with the other scaling factors listed above, but with a composition consisting only of H$_2$O. The altitudes are chosen to correspond approximately to nightside spectral observing windows in the near-infrared and to cover dynamically distinct regions of the atmosphere (shown in Figure \ref{fig:age_of_air}). Our study does not examine whether it is physically plausible for buoyant plumes from volcanic eruptions to rise to these heights. Rather, we test whether such plumes, if they exist, would be detectable and distinguishable from the background environment given the expected wind velocities and internal variability of the Venus atmosphere. We duplicate the plumes at equatorial and high latitudes to investigate the effect of different wind velocities, in particular, the decline in wind speeds polewards of 60\textdegree ~in both hemispheres. The longitudes of the plumes have a negligible effect on their dispersion, but have been chosen to correspond roughly to the locations of Maat Mons and Skadi Mons as the highest elevation points at these latitudes.

The model is configured to output data at a very high cadence of four times per Earth hour to capture the rapid development of the plumes. To keep the output size at a manageable level, and because initial tests showed the plume dispersal is rapid, each simulation is run for 0.1 Venus solar days (about 12 Earth days). The first 28-hour period of this runtime represents an active eruption: the enhanced volume mixing ratio in the gridbox is held constant and the winds can advect the volcanic gases downstream. The enhancement is then stopped and the plume is allowed to disperse freely. The injection time of 28 hours was chosen to approximate the longest known explosive phases of eruptions on Earth. The 2011 Cord\'on Caulle eruption in Chile lasted for about one day \parencite{bonadonna_dynamics_2015} and the 1912 Novarupta eruptions in Alaska consisted of three distinct explosive phases covering approximately three days \parencite{fierstein_plinian_1992}. Many Plinian-style explosive eruptions have shorter timescales, including Pinatubo at nine hours \parencite{holasek_satellite_1996} and Hunga Tonga with multiple shorter explosive phases of a few hours each \parencite{vaneaton_lightning_2023a}. Since the timescales of explosive eruptions on Venus, if they exist, are unknown, we performed a second set of sensitivity simulations varying the plume injection time with values of approximately 6, 12, 38, and 126 hours. The simulation time of the longer plume injection tests was increased to 0.3 or 0.5 Venus solar days to ensure the full plume dispersal was captured.

On the gridbox level, the plume is considered dispersed when the trace gas abundances have returned to the values measured in the plume gridbox immediately before the eruption. As the gases in each gridbox have some intrinsic variability, the value is considered returned to baseline when it is within a tolerance of the initial gridbox value. The dispersal time $\tau_d$ is therefore the interval between the end of the plume forcing and the time when the gridbox value has returned to the baseline before the eruption, plus the specified tolerance:

\begin{equation}
\tau_d = t_f - t_0
\end{equation}

where $t_f$ is the time when the gridbox value has returned to the value before the eruption, allowing for the tolerance, and $t_0$ is the time when the plume forcing ceases. For individual gridboxes, the tolerance is 0.5\%. For the global maps of dispersal shown in Section \ref{sec:results}, the baseline is taken to be the global long-term mean and the tolerance is increased to 5\% to allow for spatial variability in the initial gridbox values, including equator-to-pole gradients. All figures and tables state the baseline and tolerance in their captions. Note that the observability of plumes is also determined by instrument sensitivity at each wavelength, so the dispersal time is not necessarily equivalent to the window of opportunity for viewing a plume.  

\section{Results} \label{sec:results}
\subsection{Vertical structure of the atmosphere} \label{subsec:results_vert_structure}
\noindent\begin{minipage}{\linewidth}
    \makeatletter
    \def\@captype{figure} 
    \makeatother
    \centering
    \includegraphics[width=10cm]{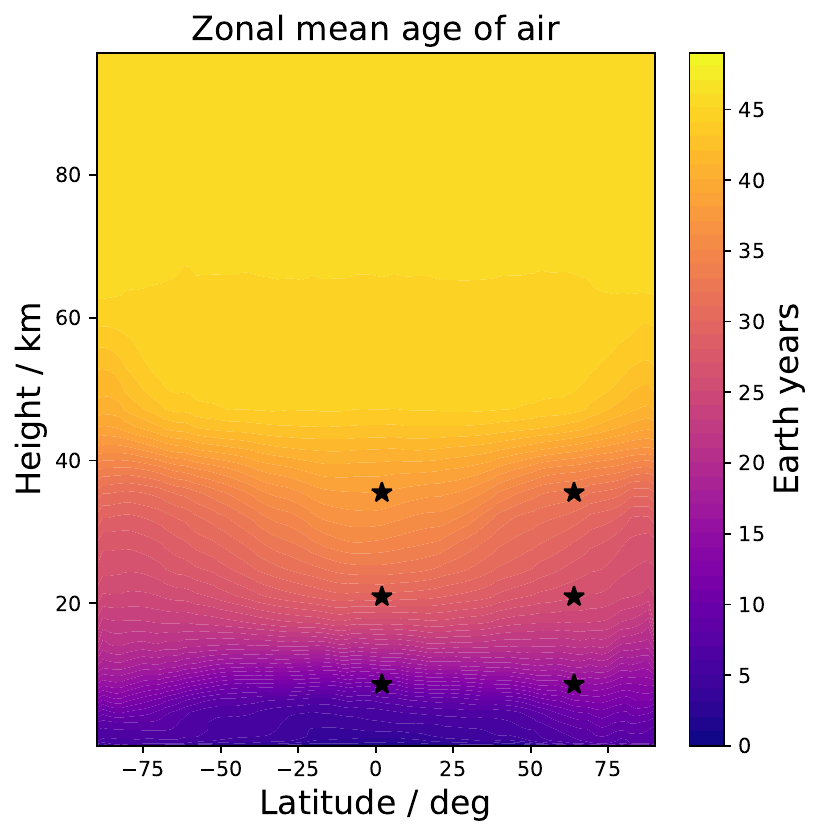}
    \caption{Transport times from the planetary surface to the model top at 98 km, zonally averaged and expressed in Earth years. The stars show the positions in latitude and altitude of the volcanic plume injections.}
    \label{fig:age_of_air}
\end{minipage}
\vspace{1em} 

Figure \ref{fig:age_of_air} shows the time for an air parcel to travel from the surface to higher altitudes. The relatively short transport timescales of months to a few years below 10 km are due to a weak, thermally direct overturning circulation caused by the small amount of solar radiation absorbed at the surface. Above this circulation at about 10 km, upward transport slows: it takes nearly three decades for an air parcel to travel from 10 km to 45 km. A latitudinal gradient appears in this region, with more rapid transport at the poles and slower movement at the equator. This gradient is caused by planetary-scale waves which form at around 50\textdegree ~N/S at 20 km and move polewards with increasing altitude, as described in detail in \textcite{cohen_planetary_2024} and \textcite{cohen_three_2025}. The waves cause vertical mixing not present at the equator. At around 47 km altitude, the rate of transport increases again until the model top. 

For the purposes of this study, we define three regions in the vertical structure of the atmosphere: 1) a near-surface "trap" which distributes tracers polewards rather than upwards below 10 km, 2) the slow-moving region between 10 and 45 km where mixing is mainly wave-driven at high latitudes, and 3) the rapidly-moving convective cloud deck at 47 km and upwards. Fig. \ref{fig:age_of_air} also shows the locations of the plumes listed in Table \ref{tab:plumes}. Plumes 1 and 2 are placed in the near-surface trap, while plumes 3, 4, 5, and 6 are placed at two altitudes in the slow-moving middle region. These altitudes were chosen because they fall into observing windows in the near-infrared part of the spectrum targeted by VenSpec-H. The two latitudes were chosen to cover the equatorial jet, where zonal and meridional winds are relatively consistent from the equator to 40 or 50\textdegree~ N/S, and the extra-jet region characterised by a drop-off in wind speeds and the presence of planetary-scale waves.

The long timescales below the cloud decks make it unlikely that near-surface emissions from volcanic eruptions can travel upwards far enough to reach more observable parts of the atmosphere, such as the cloud tops or the spectral window at 35 km. For example, even if numerous small or non-explosive eruptions were to deposit large fluxes of volcanic gases in the bottom 10 km of the atmosphere, these would become diluted or undergo chemical reactions in the 30-35 years it would take to reach 35 km. If a plume-like increase in volcanic gases is detected in the Venus troposphere, it is therefore likely to have been deposited directly at the relevant altitude rather than advected vertically from elsewhere, or to represent some other form of variability than a volcanic eruption. 

To give an idea of how a plume interacts with these wind fields, Figure \ref{fig:plume_xsect} depicts a streak created by a plume injection (Plume 6 in Table \ref{tab:plumes}) as it develops over time. The plume rapidly diffuses westwards in longitude with the background zonal wind, but generates a much smaller cross-section in latitude and altitude. This is expected given the primarily zonal flow of Venus and the slow upward transport times established in Fig. \ref{fig:age_of_air}.

\noindent\begin{minipage}{\linewidth}
    \makeatletter
    \def\@captype{figure} 
    \makeatother
    \centering
    \includegraphics[width=15cm]{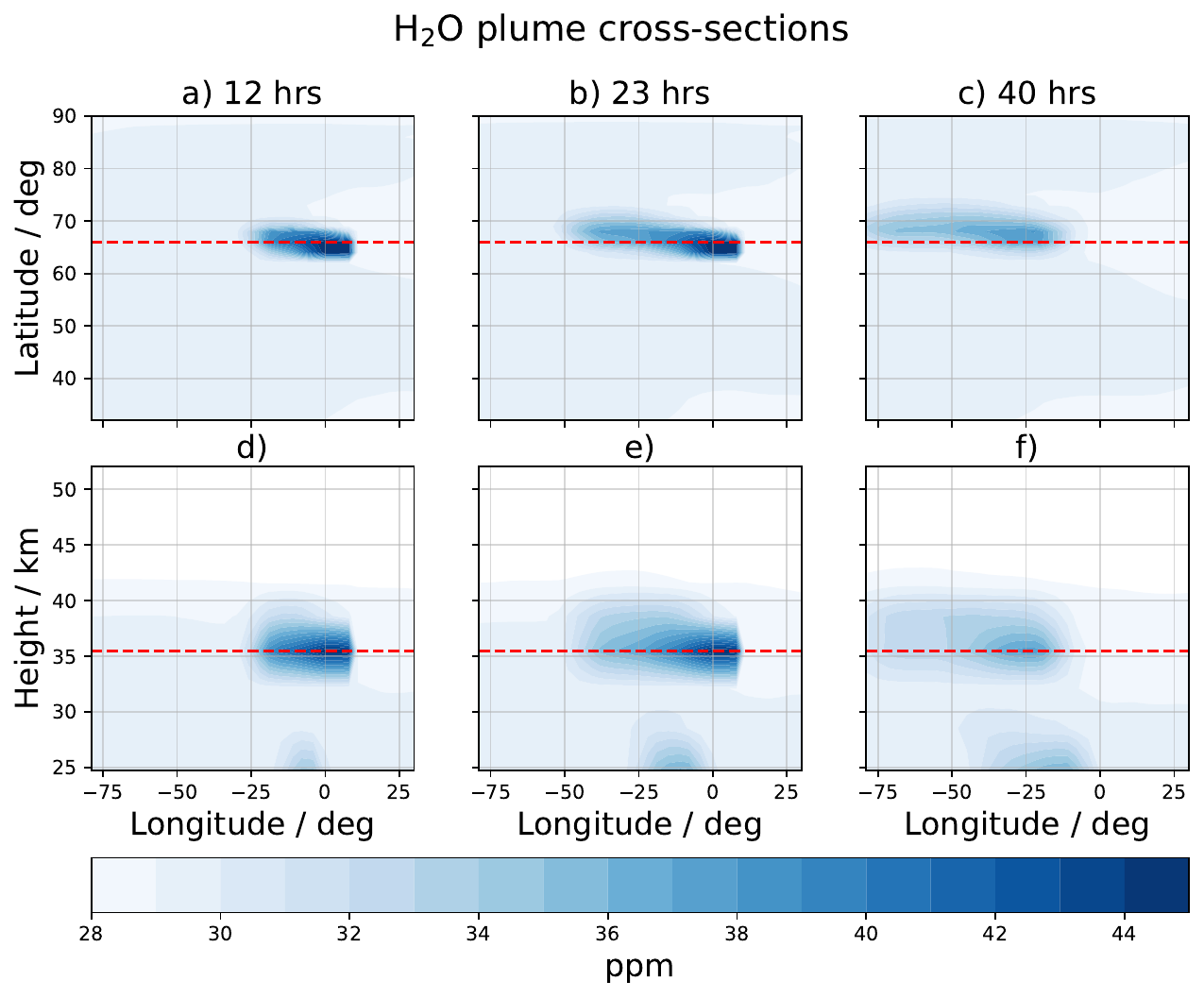}
    \caption{Longitude-latitude (top row) and longitude-height (bottom row) cross-sections of a high-latitude water vapour plume at 35 km (Plume 6 in Table \ref{tab:plumes}) at three different times. The red dashed lines in the top row show where the altitude cross-section in the bottom row is taken, while the red dashed lines in the bottom row show the latitude cross-section in the top row.}
    \label{fig:plume_xsect}
\end{minipage}
\vspace{1em} 

\subsection{Plumes injected at 8.62 km} \label{subsec:results_plumes_8.62km}
\begin{figure}[h]
    \centering
    \includegraphics[width=10cm]{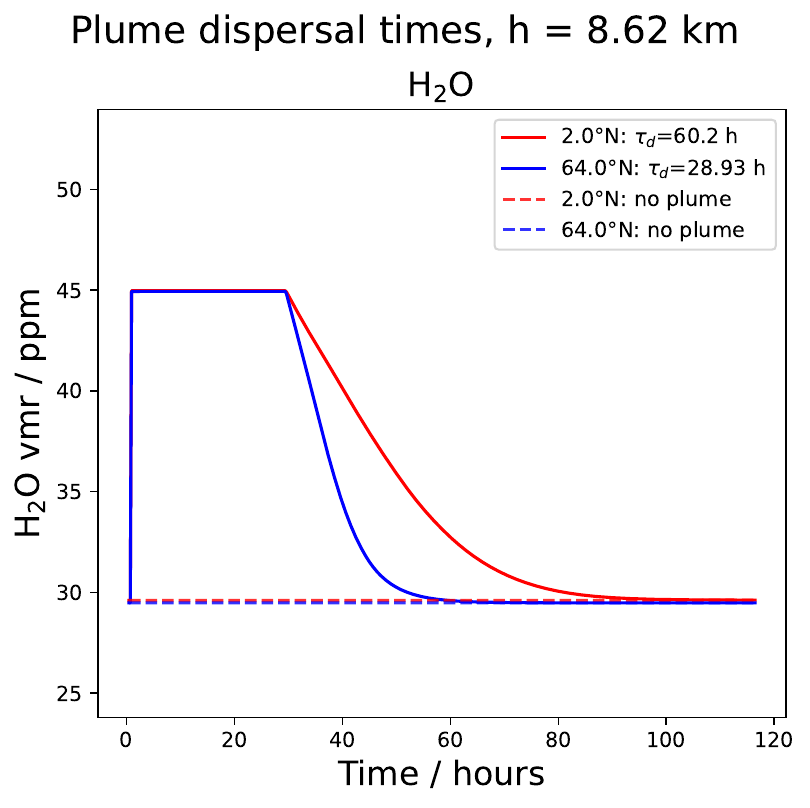}
    \caption{Dispersal time for plumes 1 and 2 measured at the coordinates given in Table \ref{tab:plumes}. The dispersal time is calculated from the time when the plume forcing stops and does not include the initial forcing period. The baseline for each plume, represented by the dashed red and blue lines, is the value in an identical simulation without the plume injection and the tolerance is 0.5\%.}
    \label{fig:dispersal_time_1_and_2}
\end{figure}

Figure \ref{fig:dispersal_time_1_and_2} shows the water vapour abundance in the plume gridbox from the start of the eruption until it returns to the background value, determined for this figure from an identical simulation run without plume injections. For the first 28 hours, the H$_2$O value is prescribed at 45 ppm. Once the "eruption" stops, the value returns to background due to dilution and dispersion by the general circulation. The model contains no chemical sinks of H$_2$O at these altitudes. The equatorial plume remains above background for 2.5 Earth days, while the high-latitude plume disperses more rapidly in about one day. However, these values are calculated only for the original plume source gridbox itself. Since the volcanic gases are advected downstream of the eruption, the area of enhancement may be both larger in size and longer lasting than at the source. Figure \ref{fig:dispersal_map_1_and_2} demonstrates this effect. A substantial area downstream of the high-latitude eruption remains 5\% above background for 75-100 hours (3-4 Earth days). During the eruption, prograde (westward-flowing) advection forms a streak of volcanic gas, which maintains a relatively coherent shape as it travels in the zonal direction due to low meridional winds and little divergence in the flow. The equatorial eruption forms a similar streak. 

For these streaks to be identifiable as plumes, the enhancement must be larger than the intrinsic background variability of H$_2$O at 8.62 km. Figure \ref{fig:variability_8.62km} shows the mean volume mixing ratio and variability of water vapour as a function of latitude and longitude at 8.62 km. The background variability is less than 0.2\% at all longitudes and latitudes. No chemical reactions are known to affect water vapour at these altitudes and the simulation does not predict dynamically-driven variability. A temporary streak-shaped local increase in H$_2$O at 8.62 km could plausibly be the result of a volcanic plume.
\clearpage
\vspace{1em}
\noindent\begin{minipage}{\linewidth}
    \makeatletter
    \def\@captype{figure} 
    \makeatother
    \centering
    \includegraphics[width=10cm]{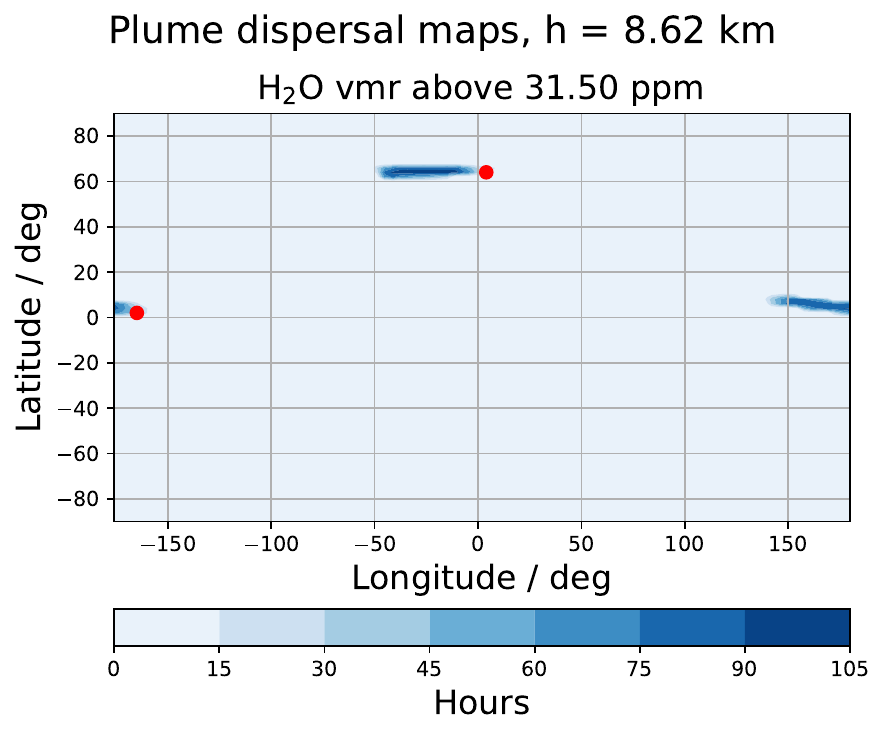}
    \caption{Dispersal time for plumes 1 and 2 for each longitude-latitude gridbox in Earth hours. The dispersal time is calculated from the time when the plume forcing stops and does not include the initial forcing period. The baseline is the global long-term mean value and the tolerance is 5\%. The red dots are plume injection sites.}
    \label{fig:dispersal_map_1_and_2}
\vspace{1em}
    \centering
    \includegraphics[width=15cm]{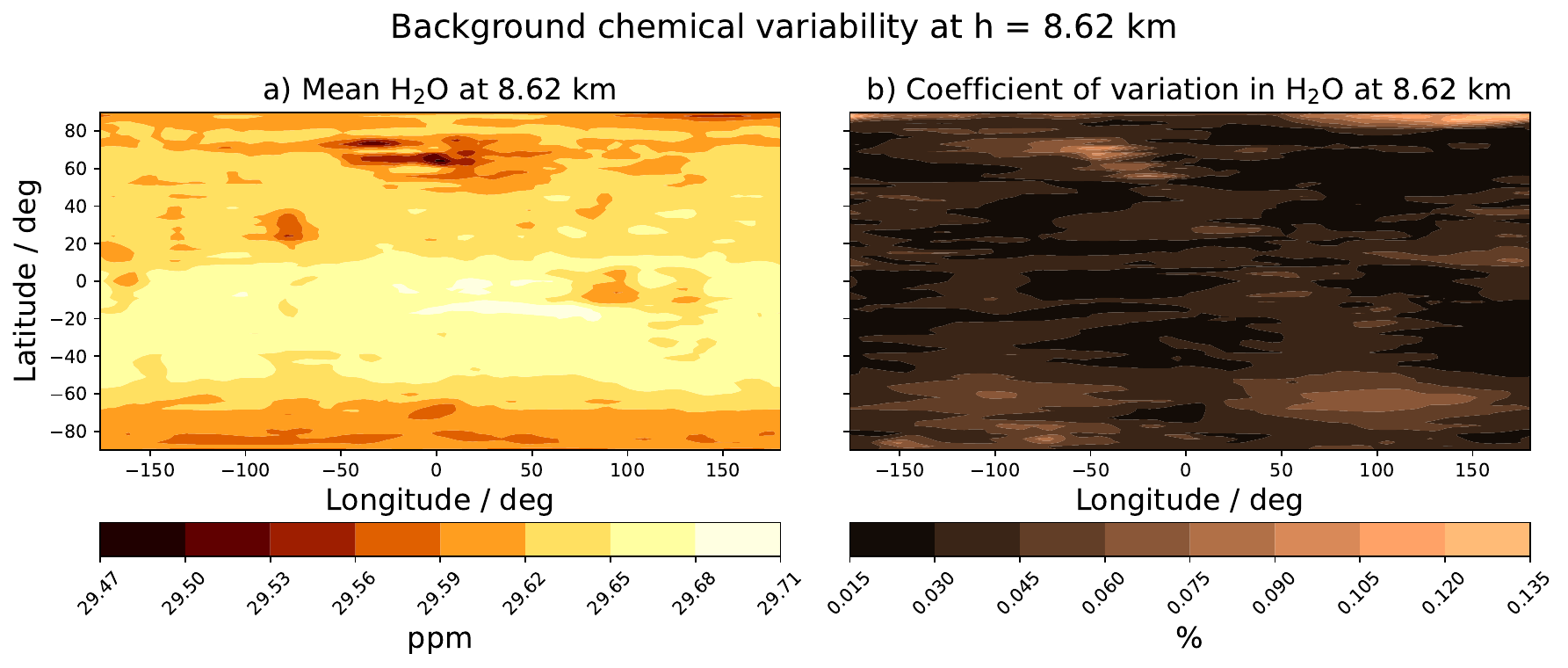}
    \caption{Latitude-longitude maps of mean water vapour volume mixing ratio and coefficient of variation (standard deviation divided by the mean) in percentage form at 8.62 km altitude. The values are calculated over four Venus solar days (468 Earth days).}
    \label{fig:variability_8.62km}
\end{minipage}
\vspace{1em} 

\subsection{Plumes injected at 20.91 km} \label{subsec:results_plumes_20.91km}
\begin{figure}[h]
    \centering
    \includegraphics[width=15cm]{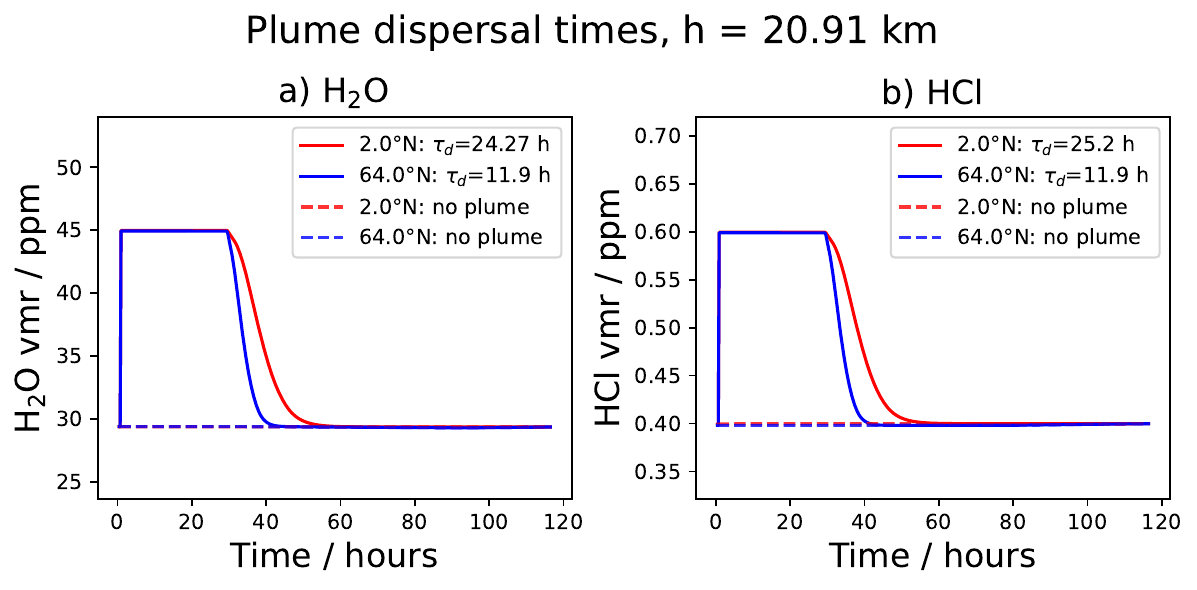}
    \caption{Dispersal time for plumes 3 and 4 measured at the coordinates given in Table \ref{tab:plumes}. The dispersal time is calculated from the time when the plume forcing stops and does not include the initial forcing period. The baseline for each plume, represented by the dashed red and blue lines, is the value in an identical simulation without the plume injection and the tolerance is 0.5\%.}
    \label{fig:dispersal_time_3_and_4}
\end{figure}

Figure \ref{fig:dispersal_time_3_and_4} shows the dispersal time for the plumes injected at 20.91 km altitude. The plume composition consists of H$_2$O and HCl, which both have spectral windows around 1.74 \textmu m. A comparison of Fig. \ref{fig:dispersal_time_3_and_4} with \ref{fig:dispersal_time_1_and_2} reveals similarities and differences. The plumes injected at 20.91 km disperse more rapidly than those at 8.62 km, as is expected given the higher wind speeds at the higher altitude. The equatorial plume again has a longer lifetime than the high-latitude plume due to lower meridional wind speeds. The dispersal times for the H$_2$O and HCl are nearly identical. Figure \ref{fig:dispersal_map_3_and_4} confirms the dispersal time and shape of the plume streak are the same in H$_2$O and HCl. This is unsurprising since both species behave as passive tracers with no active chemistry at these model altitudes. The dispersal time for a plume at 20.91 km is a brief 0.5-1 Earth days, depending on latitude, after the eruption ends at its source, but Fig. \ref{fig:dispersal_map_3_and_4} shows that the trace gas abundances remain enhanced by at least 5\% for up to 2.5 days westward of the volcano.

The background variability in H$_2$O and HCl remains extremely low at this altitude. Figure \ref{fig:variability_20.91km} shows that H$_2$O varies by a maximum of 0.15\% and HCl by 0.014\% -- an order of magnitude less. Even an eruption that causes an increase of just 1\% should be distinguishable from the background, especially at low latitudes. Volcanic eruptions on Earth easily meet this threshold, causing HCl enhancements of up to 9 times the background value of approximately 1 ppb (\textcite{carn_multi-decadal_2016}, Table 7). While the dispersal time is less at 20.91 km than 8.62 km, the potential to measure two volcanic gas species abundances could increase certainty in a plume detection if both species display a temporary zonal streak structure that could be downstream of an eruption.

\noindent\begin{minipage}{\linewidth}
    \makeatletter
    \def\@captype{figure} 
    \makeatother
    \centering
    \includegraphics[width=15cm]{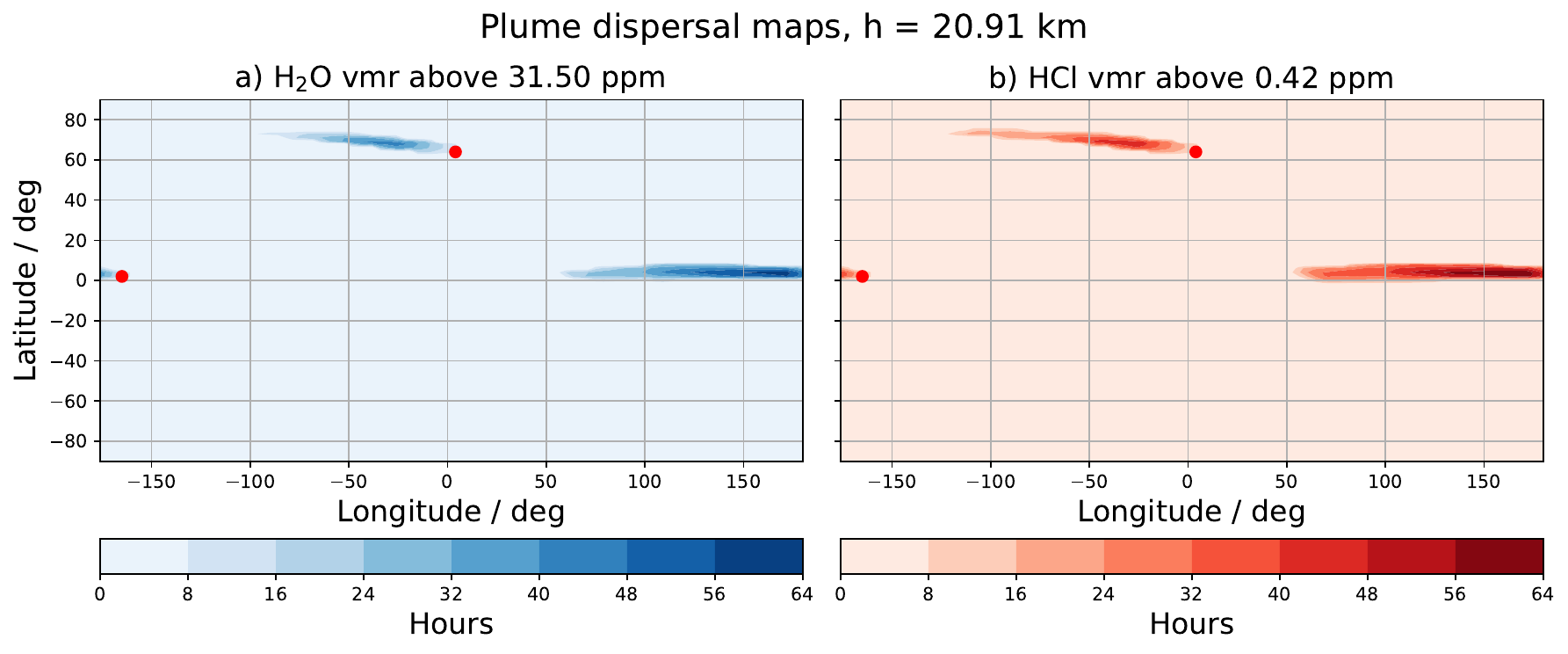}
    \caption{Dispersal time for plumes 3 and 4 for each longitude-latitude gridbox in Earth hours. The dispersal time is calculated from the time when the plume forcing stops and does not include the initial forcing period. The baseline is the global long-term mean value and the tolerance is 5\%.  The red dots are plume injection sites.}
    \label{fig:dispersal_map_3_and_4}
\vspace{1em}
    \centering
    \includegraphics[width=15cm]{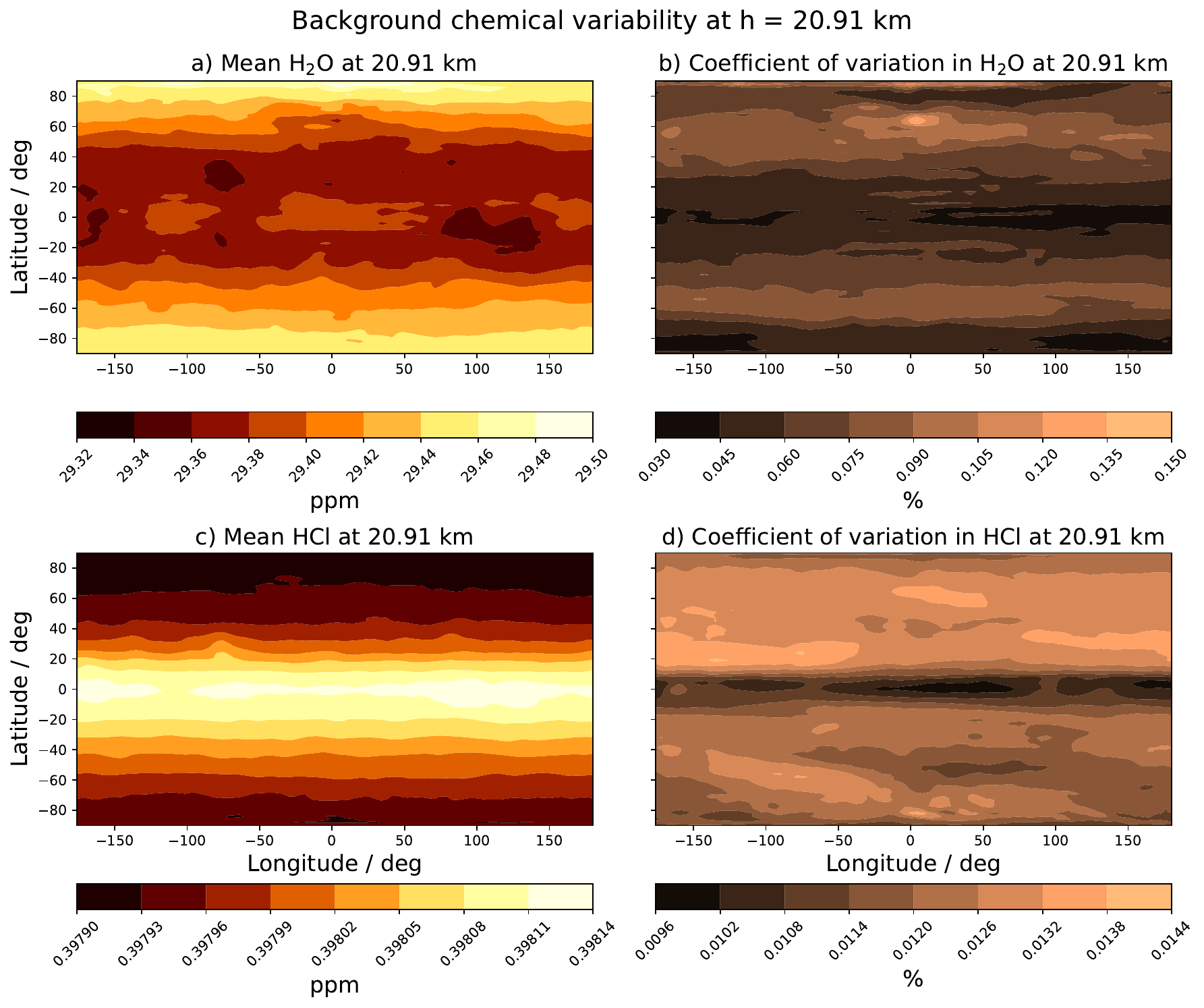}
    \caption{Latitude-longitude maps of mean water vapour and hydrogen chloride volume mixing ratios and coefficient of variation (standard deviation divided by the mean) in percentage form at 20.91 km altitude. The values are calculated over four Venus solar days (468 Earth days).}
    \label{fig:variability_20.91km}
\end{minipage}
\vspace{1em} %

\subsection{Plumes injected at 35.45 km} \label{subsec:results_plumes_35.45km}
\begin{figure}[h]
    \centering
    \includegraphics[width=10cm]{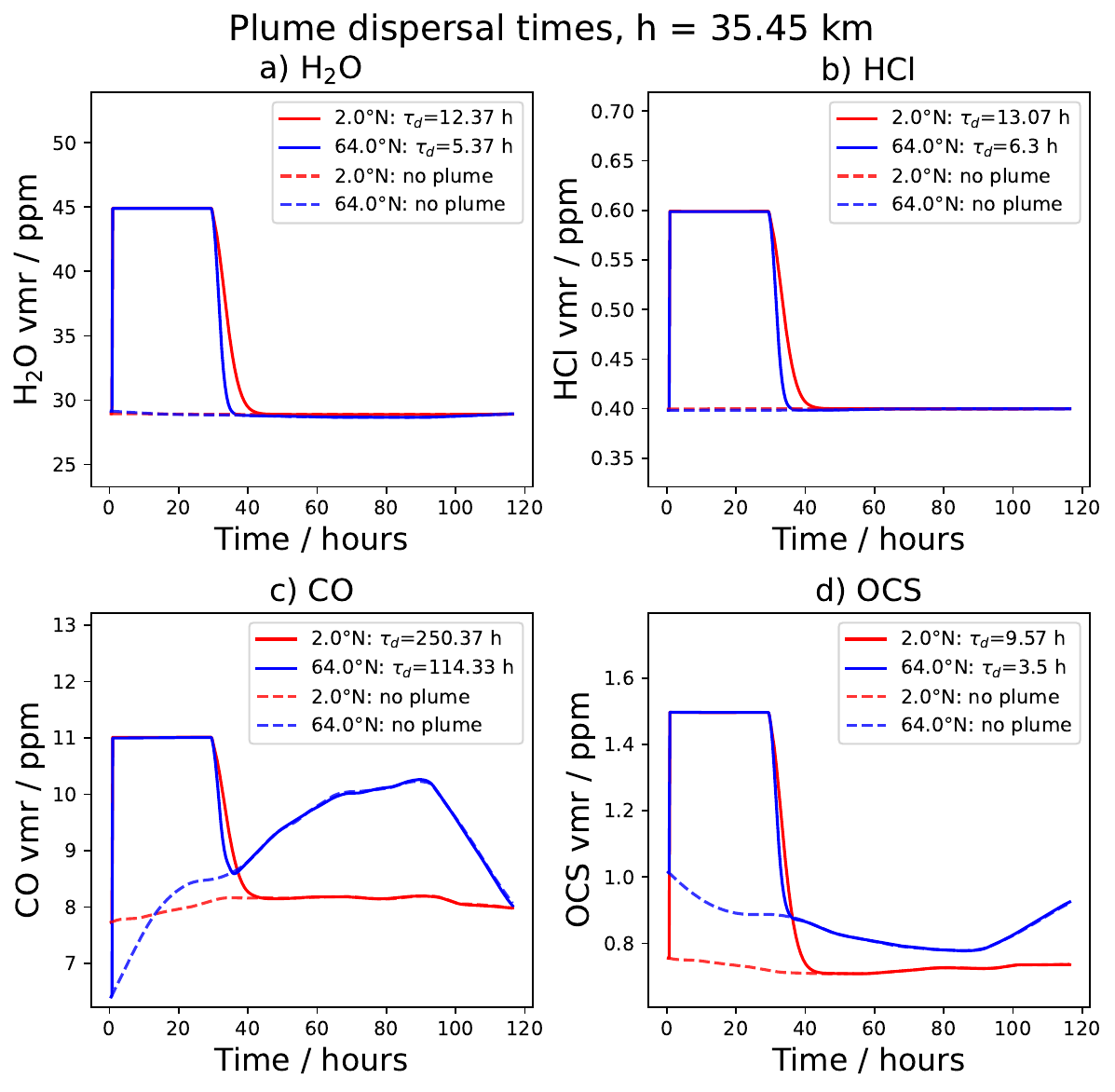}
    \caption{Dispersal time for plumes 5 and 6 measured at the coordinates given in Table \ref{tab:plumes}. The dispersal time is calculated from the time when the plume forcing stops and does not include the initial forcing period. The baseline for each plume, represented by the dashed red and blue lines, is the value in an identical simulation without the plume injection and the tolerance is 0.5\%.}
    \label{fig:dispersal_time_5_and_6}
\end{figure}

The plumes injected at 35.45 km display different behaviour than those at lower altitudes. These plumes consist of both inert species (H$_2$O, HCl) and chemically active species (CO, OCS). The dispersal times of H$_2$O, HCl, and OCS for equatorial and high-latitude plumes shown in Figure \ref{fig:dispersal_time_5_and_6} follow a similar pattern as the lower-altitude plumes: a rapid drop in volume mixing ratio in the source gridbox as soon as the eruption stops. For H$_2$O and HCl, the equatorial plume disperses in around 12 hours and the high-latitude plume in 5-6 hours--twice as quickly as the plumes at 20.91 km. This can be attributed to the faster wind speeds at 35.45 km advecting trace gases out of the source gridbox more rapidly than at the lower altitude. However, the values for CO and OCS take much longer to return to the value before the plume injection than the H$_2$O and HCl at the same altitude. Comparison values from an identical simulation without plumes show the background variability of CO and OCS at the time of the plume injection. The high-latitude CO abundance (blue line in Fig. \ref{fig:dispersal_time_5_and_6}c) begins to drop, then increases before dropping again around 100 hours later in essentially a second peak after the eruption. The high-latitude OCS abundance (blue line in Fig. \ref{fig:dispersal_time_5_and_6}d) undergoes a swift decline and drops below the initial value for hundreds of hours before it begins to rise again after 100 hours. The intrinsic variability of these species makes it difficult to define when a plume has vanished or to distinguish the plume from background variability, even though comparison to a simulation without plumes reveals that the plume enhancement disappears almost immediately.

\begin{figure}[h]
    \centering
    \includegraphics[width=15cm]{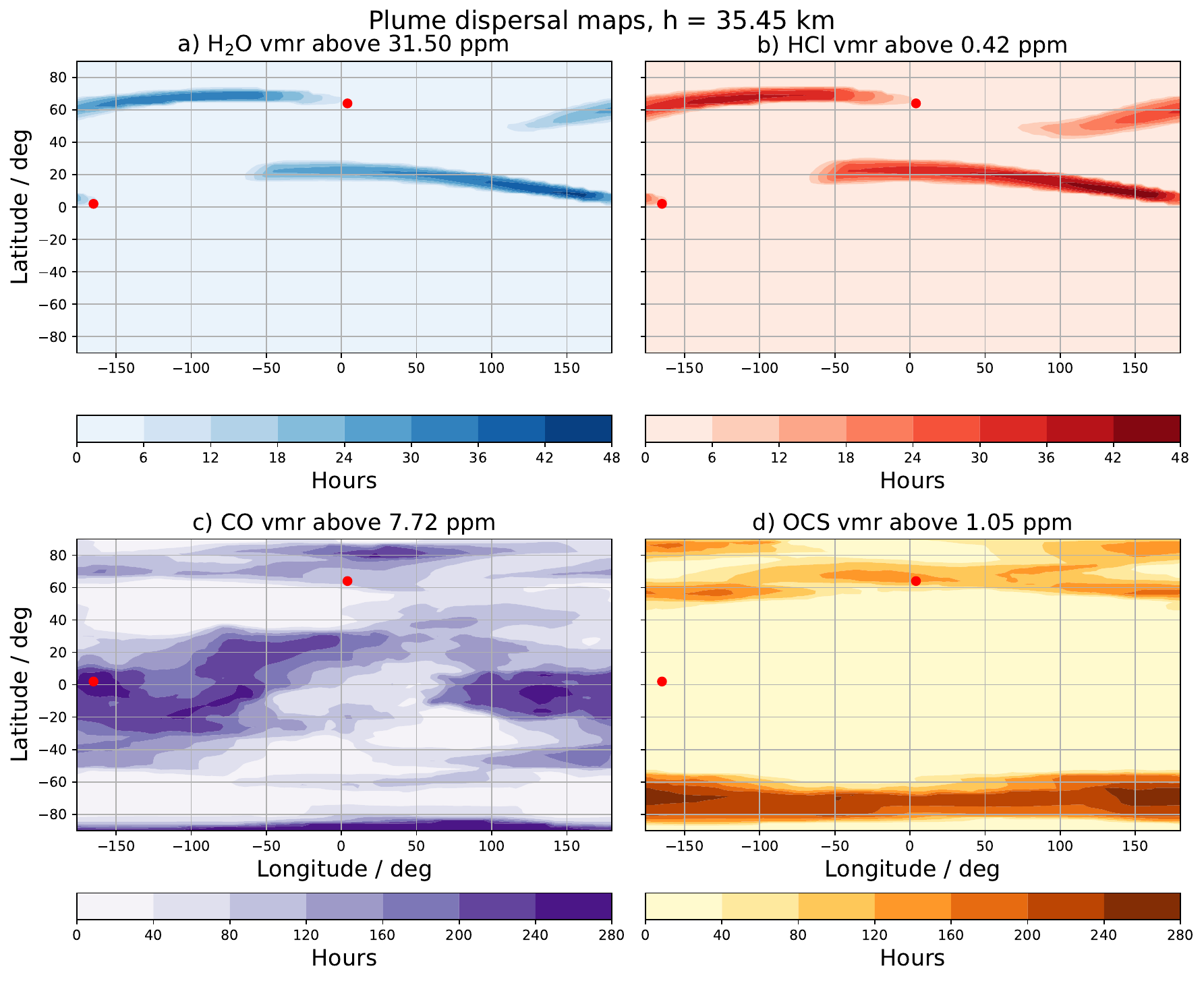}
    \caption{Dispersal time for plumes 5 and 6 for each longitude-latitude gridbox in Earth hours. The dispersal time is calculated from the time when the plume forcing stops and does not include the initial forcing period. The baseline is the global long-term mean value and the tolerance is 5\%.  The red dots are plume injection sites.}
    \label{fig:dispersal_map_5_and_6}
\end{figure}

Figure \ref{fig:dispersal_map_5_and_6} maps the dispersal times for each plume in longitude and latitude. The maps for H$_2$O and HCl form long streaks downstream of the eruptions with a peak in the period of enhancement (of 48 hours) west of the eruption site, as seen at the lower altitudes. However, the maps for CO and OCS show a different pattern. There are no streak shapes; instead, the abundances of each species are above the baseline value for hundreds of hours in large spatial regions at high latitudes for both CO and OCS and also at equatorial latitudes for CO. Instead of plumes, the dispersal time metric detects the spatial and temporal background variability of these two species.

\clearpage
\begin{figure}[h]
    \centering
    \includegraphics[width=12cm]{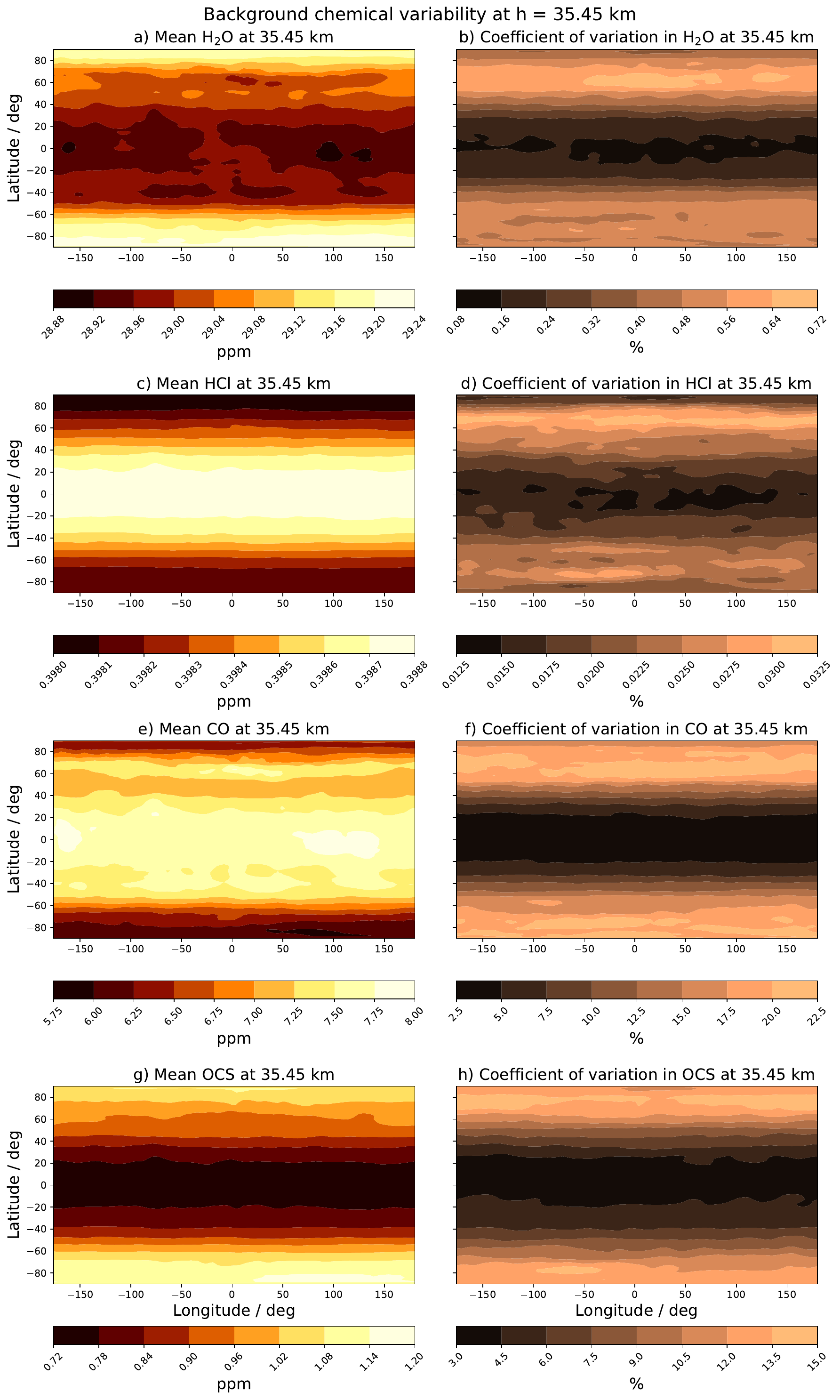}
    \caption{Latitude-longitude maps of mean water vapour, hydrogen chloride, carbon monoxide, and carbonyl sulphide volume mixing ratios and coefficient of variation (standard deviation divided by the mean) in percentage form at 35.45 km altitude. The values are calculated over four Venus solar days (468 Earth days).}
    \label{fig:variability_35.45km}
\end{figure}
\clearpage

\begin{figure}[h]
    \centering
    \includegraphics[width=15cm]{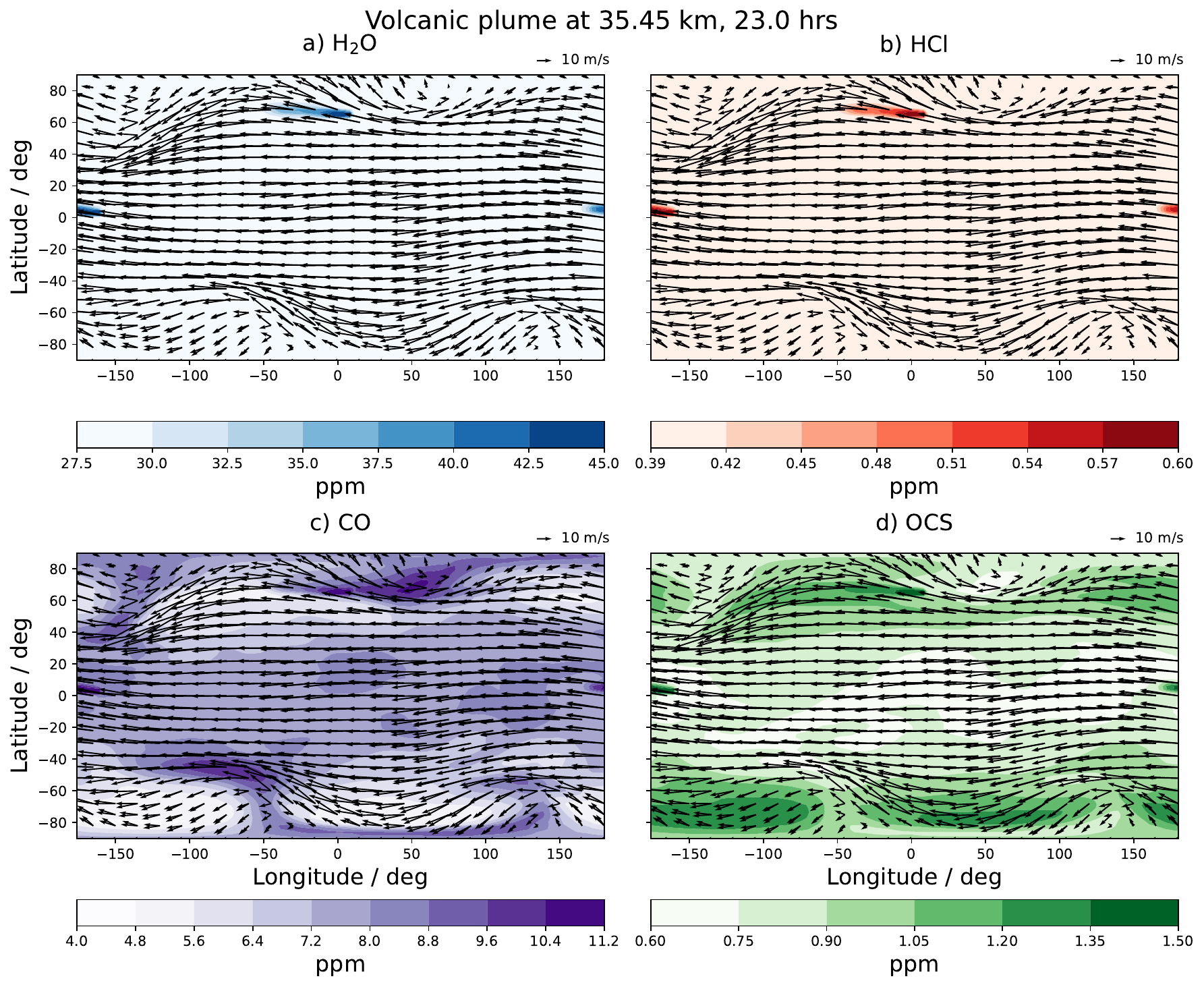}
    \caption{Longitude-latitude snapshot of water vapour, hydrogen chloride, carbon monoxide, and carbonyl sulphide volume mixing ratios at 35.45 km altitude, 23 hours after the beginning of the eruption. Arrows represent the horizontal wind vectors. The 0\textdegree ~longitude line is midnight in local time.}
    \label{fig:snapshot_35.45km}
\end{figure}

Figure \ref{fig:variability_35.45km} reveals why this occurs. The left column shows longitude-latitude maps of the mean volume mixing ratio of each species and the right column shows the corresponding coefficient of variation as a percentage. All species display greater variability at latitudes above 50\textdegree, but for H$_2$O and HCl the magnitude of variability peaks at 0.72\% and 0.0325\%, respectively, while for CO and OCS it peaks at 22.5\% and 15\%. Even the minimum variability at equatorial latitudes for the latter two species starts at 2.5-3.0\%. The greater variability in all species polewards of 50\textdegree ~can be attributed to planetary-scale waves which form at these latitudes \parencites{cohen_three_2025, cohen_planetary_2024}. However, the much larger magnitude of the variability in CO and OCS must be due to the chemically active nature of these species at this altitude. The CO and OCS maps in Fig. \ref{fig:variability_35.45km} suggest that both atmospheric dynamics and chemical reactions play a role in the variability of these species, and the spatial maps of dispersal time in Fig. \ref{fig:dispersal_map_5_and_6} indicate that the timescales of the background variability are short enough to confound detection of plume streaks. Figure \ref{fig:snapshot_35.45km}, which shows a snapshot of the eruption plume in all four trace gas species, illustrates how the high CO and OCS variability mask the plumes. (Equivalent snapshots for the two lower altitudes are provided as Figures S1 and S2 in the Supplemental Materials.) The CO abundance displays a zonal wavenumber-1 pattern at high latitudes. The maximum of this wave pattern passes over the volcano site not long after the eruption. In Fig. \ref{fig:dispersal_time_5_and_6}, the wave maximum causes a signal similar in magnitude, duration, and shape to the plume. 

\subsection{Sensitivity tests} \label{subsec:results_sensitivity}
\begin{figure}[h]
  \centering
  \includegraphics[width=15cm]{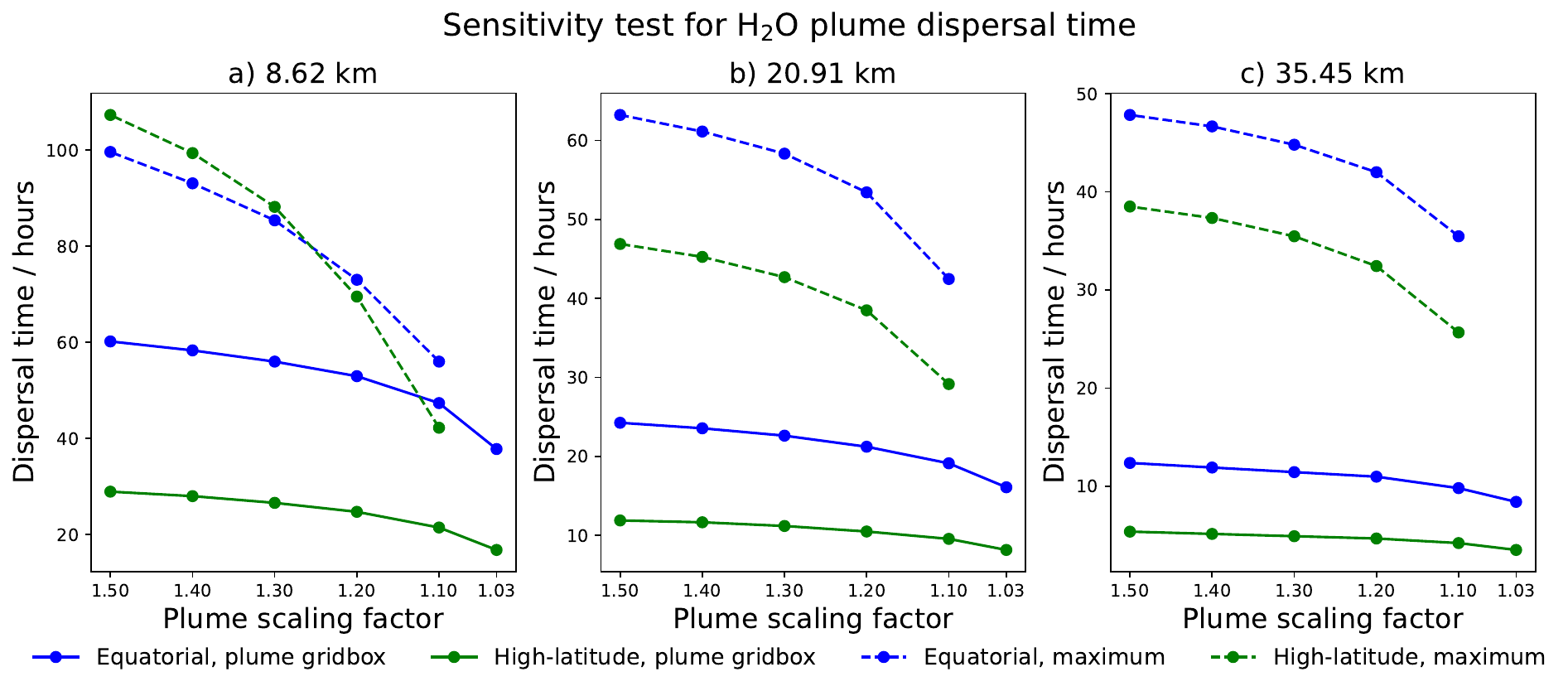}
  \caption{Dispersal time at the plume gridbox and maximum dispersal time for a plume of H$_2$O for different scaling factors and for three different altitudes. The baseline for the plume gridbox is the value in the gridbox immediately before the eruption and the tolerance is 0.5\%. The baseline for the maximum dispersal time is the global long-term mean value and the tolerance is 5\%. The 1.03 scaling factor case is omitted for the maximum dispersal time as it is below the tolerance threshold. The maximum values are found at different longitude-latitude locations as shown in Figs. \ref{fig:dispersal_map_1_and_2}, \ref{fig:dispersal_map_3_and_4}, and \ref{fig:dispersal_map_5_and_6}.}
  \label{fig:sensitivity}
\end{figure}

We scaled the trace gas abundances in our nominal plume simulation by a factor of 1.5 (50\% increase from baseline) to align with the work of \textcite{dias_volcanic_2025}. In their Figure 3 and Table 1, the H$_2$O volume mixing ratio between 30 and 40 km is taken to be 44 to 48 ppm (compared to our 45 ppm), corresponding to a 10\% increase in total integrated column density. However, as smaller eruptions are more probable, we performed sensitivity tests by repeating the simulation with scaling factors of 1.4, 1.3, 1.2, 1.1, and 1.03. Since the CO and OCS plumes cannot be reliably distinguished from background variability even for the nominal case, and the HCl dispersal times are nearly identical to those for H$_2$O below the clouds, we show in Figure \ref{fig:sensitivity} only the results for H$_2$O.

For equatorial eruptions that increase the water vapour abundance by 10\% at 8.62 km, 20.91 km, and 35.45 km, the dispersal times are reduced to $\sim$47, 19, and 10 hours, respectively. For reference, the Supplemental Materials include Table S1 which gives the values of the dispersal times shown in Fig. \ref{fig:sensitivity}. The dispersal times are calculated for the plume source gridbox, but the longitude-latitude maps in Figures \ref{fig:dispersal_map_1_and_2}, \ref{fig:dispersal_map_3_and_4}, and \ref{fig:dispersal_map_5_and_6} show that the plume-driven enhancement is visible for longest downstream of the eruption. The time before the plume becomes indistinguishable from the background will therefore be longer at downstream longitudes in each case. As expected, smaller eruptions disperse faster; however, the results are much less sensitive to the scaling factor than to latitude or altitude. This indicates that a larger eruption will not necessarily have a significantly longer-lasting plume.

Fig. \ref{fig:sensitivity} also shows the maximum values from the maps in Figs. \ref{fig:dispersal_map_1_and_2}, \ref{fig:dispersal_map_3_and_4}, and \ref{fig:dispersal_map_5_and_6} for the nominal case and all sensitivity tests except for the 1.03 scaling factor case, which is below the 5\% tolerance applied as a global threshold. The values are given for reference in Table S1 in the Supplemental Materials. In all cases, there are areas where the enhanced trace gas abundances persist longer than at the eruption site. The maximum dispersal time may be 2-3x as long as at the plume gridbox; however, even the highest value is only 100 hours (4 Earth days).

\begin{figure}[h]
  \centering
  \includegraphics[width=15cm]{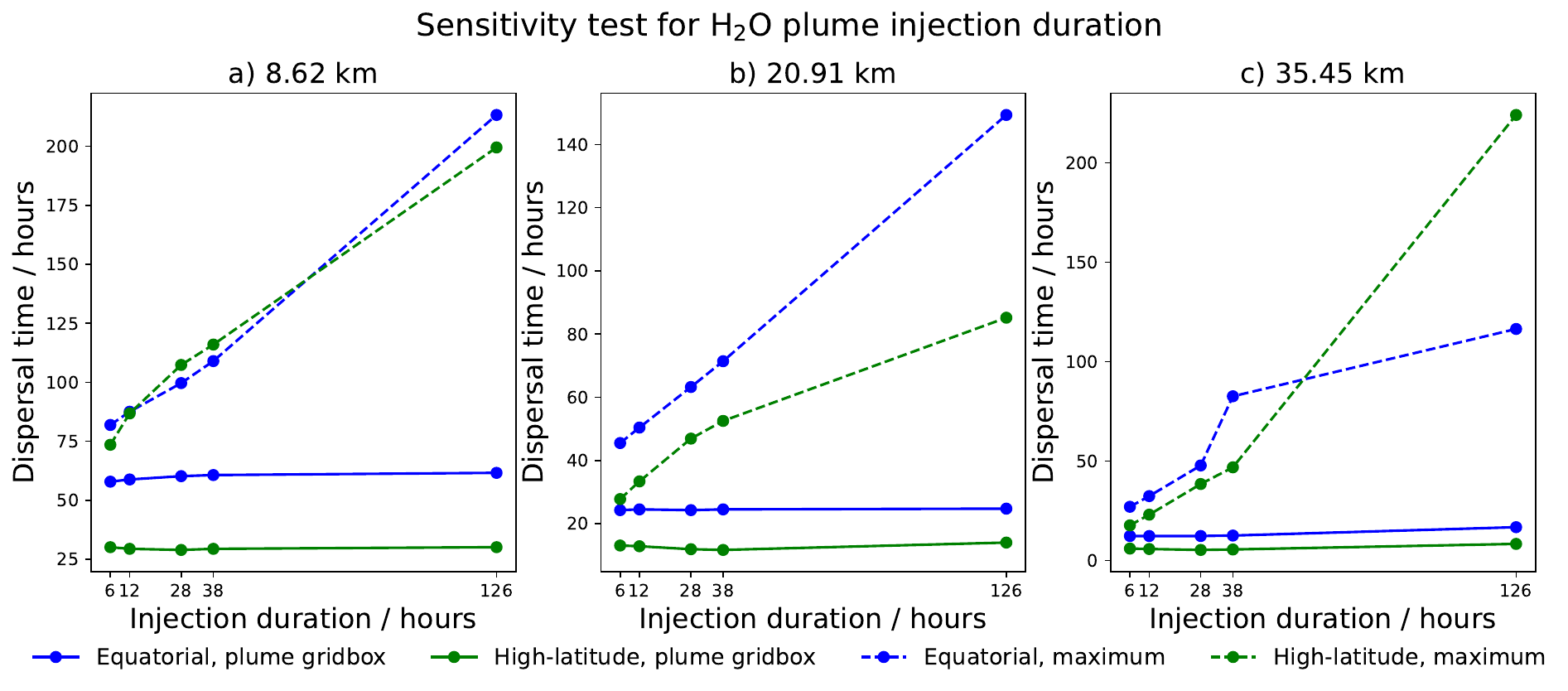}
  \caption{Dispersal time at the plume gridbox and maximum dispersal time for a plume of H$_2$O for different plume injection times and for three different altitudes. The baseline for the plume gridbox is the value in the gridbox immediately before the eruption and the tolerance is 0.5\%. The baseline for the maximum dispersal time is the global long-term mean value and the tolerance is 5\%. The maximum values are found at different longitude-latitude locations as shown in Figs. \ref{fig:dispersal_map_1_and_2}, \ref{fig:dispersal_map_3_and_4}, and \ref{fig:dispersal_map_5_and_6}.}
  \label{fig:time_test}
\end{figure}

In a second set of sensitivity tests, we held the scaling factor fixed at 1.5, but varied the plume injection time. We repeated the nominal 28-hour simulation with plume injection times of approximately 6 hours, 12 hours, 38 hours, and 126 hours. Six hours and 12 hours are more realistic based on Plinian eruptions on Earth; 28 hours is an upper limit based on Earth, with 38 and 126 hours included to show the trend if longer explosive eruptions prove to be possible on Venus. As in Fig. \ref{fig:sensitivity}, we only show the results for H$_2$O. Figure \ref{fig:time_test} displays how the dispersal time at the plume gridbox and the maximum dispersal time vary as the plume injection time increases. The figure values are also given in Table S2 in the Supplemental Materials. The plume gridbox dispersal time is nearly constant. This is because the plume gridbox empties at the same speed regardless of how long it has been forced. Increasing the plume injection gives the eruption a longer time to emit volcanic trace gases into the atmosphere, increasing the maximum dispersal time. The maximum dispersal time at 8.62~km is fairly linear, with little difference between the equatorial and high-latitude plumes. At 20.91~km, the high-latitude plume disperses more quickly due to the presence of high-latitude Rossby waves (absent at the lower altitude). At 35.45~km, the high-latitude plume appears to persist unexpectedly long. This is because the excess water vapour becomes confined in the polar region above 80\textdegree N instead of dispersing, leaving a small enhancement compared to the global mean. The enhancement does not take the shape of a downstream streak as in the other cases studied; however, it suggests the possibility that ongoing volcanic activity at high latitudes could contribute to compositional gradients across a polar vortex boundary. It is not known whether the Venus atmosphere forms a polar vortex at these altitudes below the cloud deck, and the VPCM's accuracy in modelling this region is questionable given the limitations of a longitude-latitude grid and polar wind filtering. Even this long-lasting enhancement caused by an implausibly continuous (5 Earth day) explosive eruption only reaches about 225 hours (a little over 9 days) time above baseline, leaving our conclusions from the nominal simulation unchanged. Trace gas enhancements caused by volcanic plume injections last a matter of days, not weeks or months.

\section{Discussion} \label{sec:discussion}
A key science goal of the European Space Agency's EnVision mission to Venus, scheduled to launch in 2031, is to search for volcanic plumes beneath the planet's cloud decks using targeted observations of nightside infrared spectral windows. NASA's VERITAS mission, with a similar launch date, will also have capability to search for increases in water vapour near the surface. Our simulations can guide interpretation of spatial and temporal variability detected by these missions.

Broadly, observing prospects for plumes improve closer to the equator and nearer to the surface. The two species in our simulations without active chemistry in the troposphere -- H$_2$O and HCl -- take longer to disperse and behave similarly to each other. An eruption near the equatorial surface that increases the H$_2$O vmr by 50\% would take 60 hours to disperse, long enough for multiple observations by the EnVision VenSpec-H/VenSpec-M or the VERITAS VEM instruments, as both spacecraft will have orbits of around 1.5 hours. Even a smaller eruption with an increase of 10\% would cause an enhancement persisting for 48 hours. As discussed in Section \ref{sec:intro}, buoyant plumes are also more likely to remain at low altitudes; for example, \textcite{lefevre_eruption_2025} found that 87\% of simulated plumes did not exceed 15 km altitude. In addition, it is possible that non-explosive volcanism could reach the 1.18 \textmu m window. Maat Mons, at roughly 8~km height, could emit volcanic gases directly at these altitudes. A series of smaller, sub-Plinian eruptions over a week or more, each generating a signature lasting 2 to 5 days, could lead to a larger and longer-lasting increase in trace gases. On the other hand, modelling of the impact of volcanic plumes on the spectrum of Venus by \textcite{dias_volcanic_2025} indicates that eruptions at lower altitudes cause much smaller changes in radiance and therefore require higher signal-to-noise ratios (SNRs) to detect.

The 30-40 km altitude may seem like a better prospect for observing eruptions, even if only a small proportion of plumes rise to this level. \textcite{dias_volcanic_2025} show that a 50\% increase in the H$_2$O mixing ratio against baseline in this range has a substantial impact on the radiance and requires low SNRs (25-50) to detect. However, our results in Fig. \ref{fig:dispersal_time_5_and_6} indicate that a plume umbrella would disperse in 12 hours at the equator and 6 at high latitudes. In a hypothetical scenario, if Venus experiences 40 to 120 volcanic eruptions per year \parencites{byrne_estimates_2022, van_zelst_comment_2022}, 2.5\% of these reach 45 km \parencite{lefevre_eruption_2025}, and each plume persists for 12 hours, this gives 12 hours (one visible eruption) to 36 hours (three visible eruptions) per year during which a plume is visible. It is also possible to observe CO and OCS in this altitude range, and \textcite{dias_volcanic_2025} find that these two species have the next biggest impact on radiance from the deep atmosphere. Since the VenSpec-H instrument can detect these species near simultaneously, one strategy could be to interpret concurrent increases in H$_2$O, CO, and OCS as driven by a volcanic plume. Our nominal simulation with a 50\% increase in gas abundances from baseline, however, finds that this degree of enhancement cannot be distinguished from the background variability of CO and OCS. Wave-driven variability and conversion of these two species into each other generates fluctuations which can appear very much like plumes in magnitude (Fig. \ref{fig:dispersal_time_5_and_6}) and even shape (Fig. \ref{fig:dispersal_map_5_and_6}). Smaller increases may not be noticeable against the natural background variability.

Finally, while past work has found that buoyant plumes at high latitudes can rise to higher altitudes \parencites{lefevre_eruption_2025, glaze_transport_1999}, these simulations did not represent the 3-D wind fields, in particular meridional wind and planetary-scale waves. Our modelling indicates that plumes disperse at least twice as quickly at high latitudes than at the equator at all altitudes. As discussed in the previous paragraph, since latitudes polewards of 50\textdegree ~also see the formation of planetary-scale winds from 20 km upwards, these regions additionally display significant internal variability on scales that obscure the presence of plumes. 

Overall, plumes persist the longest close to the surface, at low- to mid-latitudes, and with a composition of chemically inert or long-lived gases such as H$_2$O or HCl. A streak-shaped enhancement of H$_2$O observed at 1.18 \textmu m westward of a volcanic region and with a lifespan of up to a few days would constitute a plausible plume detection. On the other hand, spatial and temporal variability in CO and OCS abundances at 35 km is far more likely to be driven by the intrinsic variability of these species in the atmosphere than by volcanic eruptions. 

\section{Conclusion} \label{sec:conclusion}
We simulated the behaviour of volcanic plumes using the Venus Planetary Climate Model to support future observations by the EnVision and VERITAS missions. We modeled plumes as gridbox-sized localised enhancements of trace gases -- specifically H$_2$O, HCl, CO, and OCS -- injected at altitudes corresponding to nightside near-infrared spectral windows (8.62, 20.91, and 35.45 km). We investigated dispersal timescales and visibility against the backdrop of the atmosphere's intrinsic chemical and dynamical variability.

Our results demonstrate that the detectability of volcanic plumes is highly dependent on altitude, latitude, and chemical composition, as well as the plume size itself. Close to the surface (8.62 km), wind speeds are relatively low, and the background variability of chemically inert species like H$_2$O and HCl is negligible (less than 0.2\%). Volcanic injections at these altitudes persist as detectable signals for extended periods, forming coherent downstream streaks. Specifically, equatorial plumes at 8.62 km remained detectable for up to 60 hours at the eruption site and up to 100 hours westward of it, while high-latitude plumes dispersed more rapidly (approx. 29 hours) due to stronger winds. These findings suggest that searching for streak-like enhancements of H$_2$O or HCl in the near-surface atmosphere, especially westward of known volcano sites, offers the most promising strategy for identifying active volcanism.

In contrast, observing prospects at 35.45 km are compromised by high intrinsic atmospheric variability. While CO and OCS are potential volcanic indicators, our simulations reveal that their background abundances fluctuate significantly due to atmospheric dynamics and chemistry. These natural fluctuations can mimic the magnitude and shape of volcanic plumes, effectively masking even a 50\% volcanic enhancement and making confident detection difficult.

To maximize the success of plume hunting on Venus, future missions should prioritise observations of chemically stable gases in the near-surface atmosphere at low latitudes, where volcanic signals persist longest and background variability is minimal.

\section*{Open science declarations}

\textbf{Acknowledgements}.

\textbf{Author contributions.} The authors contributed to this work as follows based on the Contributor Role Taxonomy (CRediT): M. Cohen -- conceptualization, formal analysis, investigation, software, visualization, writing (original draft), writing (review \& editing). J. Holmes -- conceptualization, project administration, writing (review \& editing). J. Egan -- writing (review \& editing). S. Lewis -- writing (review \& editing), funding acquisition. M. Patel - writing (review \& editing), funding acquisition.

\textbf{Code availability.} The Venus Planetary Climate Model source code is freely available at 

\href{http://svn.lmd.jussieu.fr/Planeto/trunk/}{http://svn.lmd.jussieu.fr/Planeto/trunk/}. The code version used in this study is revision 3897. The repository is managed by the Laboratoire de Météorologie Dynamique of the Institut Pierre-Simon Laplace.
The Python code used to perform the analysis and generate the figures can be accessed on 
\href{https://github.com/maureenjcohen/make_plumes_2026}{GitHub}.

\textbf{Data availability.} The nominal simulation data is available on \href{https://doi.org/10.21954/ou.rd.31852927}{The Open University's data repository ORDO}.

\textbf{Funding.} We are grateful to the UK Science and Technology Facilities Council for funding support through STFC grant ST/X001180/1 and the UK Space Agency for funding support through grant UKRI2545 and ST/X006549/1.

\textbf{Competing interests.} The authors declare that they have no known competing financial interests or personal relationships that could have appeared to influence the work reported in this paper.

\printbibliography
\vfill

\noindent
\begin{tabular}{p{0.01\textwidth} p{0.32\textwidth} p{0.01\textwidth} p{0.43\textwidth} p{0.1\textwidth}}
\cline{1-5}
\scriptsize{\textcopyright}
& \scriptsize{The Author(s) 2026.}
&
& \multirow{2}{*}{\parbox{0.43\textwidth}{\vspace{1mm}\scriptsize{This work is licensed under a \href{https://creativecommons.org/licenses/by/4.0/}{Creative Commons Attribution License 4.0} which permits unrestricted use, distribution, and reproduction in any medium, provided the original work is properly cited.}}}
& \vspace{1mm}\includegraphics[height=1.5em]{assets/cc.png}\hspace{0.15em}\includegraphics[height=1.5em]{assets/by.png}
\end{tabular}

\end{document}